\documentclass[english]{sbrt}
\usepackage[english]{babel}
\usepackage[utf8]{inputenc}

\usepackage{acro}
\DeclareAcronym{3GPP}{
    short = 3GPP,
    long  = Third Generation Partnership Project
}
\DeclareAcronym{5G}{
    short = 5G,
    long  = fifth generation
}
\DeclareAcronym{5GC}{
    short = 5GC,
    long  = 5G Core
}  
\DeclareAcronym{ADA}{
	short = ADA,
	long  = AdaBoost
}
\DeclareAcronym{AP}{
	short = AP,
	long  = access point
}
\DeclareAcronym{AWGN}{
	short = AWGN,
	long  = additive white gaussian noise
}
\DeclareAcronym{CCP}{
short = CCP,
long  = convex-concave procedure
}
\DeclareAcronym{CDF}{
    short = CDF,
    long  = cumulative distribution function
}
\DeclareAcronym{CLI}{
short = CLI,
long  = cross-link interference
}
\DeclareAcronym{CPU}{
	short = CPU,
	long  = central processing unit
}
\DeclareAcronym{CVXPY}{
	short = CVXPY,
	long  = Convex Optimization in Python
}
\DeclareAcronym{D-TDD}{
	short = D-TDD,
	long  = dynamic time division duplex
}
\DeclareAcronym{D-MIMO}{
short = D-MIMO,
long  = distributed multiple input multiple output
}
\DeclareAcronym{DL}{
	short = DL,
	long  = downlink
}
\DeclareAcronym{FP}{
	short = FP,
	long  = fractional programming
}  
\DeclareAcronym{GBR}{
    short = GBR,
    long  = Gradient Boosting Regressor
}
\DeclareAcronym{gNB}{
    short = gNB,
    long  = next generation NodeB
}
\DeclareAcronym{ISAC}{
short = ISAC,
long  = integrated sensing and communication
}
\DeclareAcronym{ITU}{
short = ITU,
long  = International Telecommunication Union
}
\DeclareAcronym{KNN}{
    short = KNN,
    long  = K-Nearest Neighbors
}
\DeclareAcronym{LoS}{
    short = LoS,
    long  = line-of-sight
}  
\DeclareAcronym{MAE}{
short = MAE,
long  = mean absolute error
}
\DeclareAcronym{MICP}{
short = MICP,
long  = mixed-integer convex optimization problem
}
\DeclareAcronym{MIMO}{
	short = MIMO,
	long  = multiple input multiple output
}
\DeclareAcronym{MINLP}{
short = MINLP,
long  = mixed-integer nonlinear problem
}
\DeclareAcronym{ML}{
    short = ML,
    long  = machine learning
}
\DeclareAcronym{mMIMO}{
short = mMIMO,
long  = massive multiple input multiple output
}
\DeclareAcronym{MSE}{
short = MSE,
long  = mean squared error
}  
\DeclareAcronym{NR}{
    short = NR,
    long  = New Radio
}
\DeclareAcronym{QoS}{
short = QoS,
long  = quality of service
}
\DeclareAcronym{RB}{
short = RB,
long  = resource block
}
\DeclareAcronym{RF}{
short = RF,
long  = Random Forest
}  
\DeclareAcronym{RMSE}{
short = RMSE,
long  = root mean squared error
}
\DeclareAcronym{RRM}{
short = RRM,
long  = radio resource management
}
\DeclareAcronym{RSRP}{
    short = RSRP,
    long  = reference signal received power
}
\DeclareAcronym{SE}{
short = SE,
long  = spectral efficiency
}
\DeclareAcronym{SCA}{
	short = SCA,
	long  = successive convex approximation
}
\DeclareAcronym{SCO}{
short = SCO,
long  = successive convex optimization
}
\DeclareAcronym{SDR}{
    short = SDR,
    long  = software-defined radio
}
\DeclareAcronym{SINR}{
short = SINR,
long  = signal to interference-plus-noise ratio
}
\DeclareAcronym{SNR}{
short = SNR,
long  = signal to noise ratio
}
\DeclareAcronym{TDD}{
	short = TDD,
	long  = time division duplex
}
\DeclareAcronym{UE}{
	short = UE,
	long  = user equipment
}
\DeclareAcronym{UL}{
short = UL,
long  = uplink
}
\DeclareAcronym{USB}{
    short = USB,
    long  = Universal Serial Bus
}
\usepackage{graphicx}
\usepackage{multirow}

\newcommand{\FigRef}[2][]{Fig.#1~\ref{#2}}

\newcommand{\TabRef}[2][]{Table#1~\ref{#2}}

\usepackage{amssymb} 

\usepackage{booktabs} 

\usepackage{subcaption}
\usepackage{float}

\usepackage[%
style=ieee,
backend=biber, 
]{biblatex}

\defbibheading{bibliography}[\refname]{\section*{#1}}

\makeatletter
\g@addto@macro{\UrlBreaks}{\UrlOrds}
\makeatother

\usepackage{xcolor}
\usepackage{tikz}
\usepackage{pgfplots}
\pgfplotsset{compat=1.18}

\begin{document}
\bstctlcite{IEEEexample:BSTcontrol}
\title{Data-Driven Case Study of gNB Placement Optimization in a Private Indoor 5G Testbed}

\author{Diogo de O. Soares, Victor F. Monteiro, Fco. Rodrigo P. Cavalcanti, Vicente A. de Sousa Jr. and J. Pedro B. Lima
\thanks{Diogo, Victor and Rodrigo are with the Wireless Telecommunications Research Group (GTEL), Federal University of Cear\'{a} (UFC), Fortaleza, Brazil (\{diogosoares, victor, rodrigo\}@gtel.ufc.br). %
Vicente is with the Leading Advanced Technologies Center of Excellence (LANCE), Universidade Federal do Rio Grande do Norte (UFRN), Natal, Brazil (vicente.sousa@ufrn.br). %
João Pedro is with Instituto Atlântico, Fortaleza, Brazil (joao\_brasil@atlantico.com.br). %
V. F. Monteiro and R. P. Cavalcanti were supported by CNPq under the  Grants DT-308267/2022-2 and DT-303625/2022-8, respectively. %
This work was partially supported by the Brazilian Funding Authority for Studies and Projects (FINEP) and the Ministry of Science, Technology and Innovation (MCTI) through the project ``AITD LITEC: Advances in Artificial Intelligence for Digital Transformation at the Laboratory of Technological Innovation and Scientific Experiments'', FINEP Reference No. 1039/24, Grant Agreement No. 01.24.0581.00. We attest that generative AI was used for linguistic editing and text refinement.}%
}

\maketitle

\markboth{XLIV BRAZILIAN SYMPOSIUM ON TELECOMMUNICATIONS AND SIGNAL PROCESSING - SBrT 2026, SEPTEMBER 29TH TO OCTOBER 2ND, 2026, SALVADOR, BA}{}

\begin{abstract}
Accurate radio planning is a fundamental requirement for the deployment of wireless networks in indoor environments, where signal propagation is strongly affected by walls, partitions, and other structural obstacles. %
Despite the availability of standardized propagation models, their ability to represent the characteristics of specific deployment scenarios is often limited, motivating the use of measurement-driven approaches. %
In this context, this paper presents a data-driven case study of \ac{gNB} placement optimization in an office using measurements collected from an experimental \ac{5G} testbed. %
A propagation model is trained from \ac{RSRP} measurements using distance and wall count as input features and integrated with a combinatorial search framework. %
The proposed workflow is used to evaluate alternative deployment strategies under different optimization criteria. %
Results indicate that satisfactory indoor coverage and improved cell-edge conditions can be achieved with a small number of \acp{gNB}. %
\end{abstract}

\begin{keywords}
    5G testbed, Indoor Radio Planning, Measurement-Based Modeling, USRP, LightGBM.
\end{keywords}

\acresetall
\section{Introduction}
The evolution of \ac{5G} mobile communications has fostered the emergence of new industrial and enterprise use cases involving real-time monitoring, automation, immersive applications and intelligent services. %
To support these applications, many organizations are adopting private \ac{5G} networks in indoor environments~\cite{Ericsson_Mobility_Indoor_2023} as an alternative to conventional wireless technologies, e.g., Wi-Fi, benefiting from dedicated spectrum access, enhanced reliability, mobility management, and fine-grained control of network resources. %
However, the successful deployment of private cellular infrastructures in indoor environments depends on accurate radio planning, since signal propagation is strongly influenced by walls, partitions, furniture, and other structural obstacles. %
As a result, the positioning of \acp{gNB} becomes a key design parameter, directly influencing coverage, service quality, and user experience. %

Several propagation models have been proposed to support indoor network planning~\cite{ITU_R_P1238_13_2025}. %
Standardized models, such as those specified by the \ac{ITU} and the \ac{3GPP}, generally estimate path loss as a function of distance while accounting for shadowing effects through statistical parameters. %
Other approaches incorporate the attenuation introduced by walls and structural obstacles through deterministic formulations. %
Although these models are widely adopted in planning tools and feasibility studies, their accuracy may be limited when applied to specific deployment scenarios whose architectural characteristics differ from the assumptions considered during model development. %

Recent advances in \ac{ML} have enabled the development of measurement-driven propagation models capable of learning complex relationships directly from empirical radio data. %
By exploiting signal measurements collected in real deployments, these approaches can capture propagation effects that are difficult to represent through analytical formulations alone. %
Nevertheless, the availability of experimental datasets remains a limiting factor, since many studies rely exclusively on simulations or publicly available traces, often lacking validation in operational wireless environments. %

Motivated by this challenge, this work investigates indoor \ac{gNB} placement optimization using measurements collected from a private \ac{5G} testbed~\cite{alves2024} deployed in a real office environment. %
The experimental infrastructure was implemented using the open-source srsRAN platform and \ac{SDR} hardware. 
An extensive measurement campaign was conducted to collect \ac{RSRP} samples under different propagation conditions. %
These measurements were then used to train a LightGBM-based propagation model using physically meaningful features derived from geometry and obstacle information. %

The resulting model is integrated with a combinatorial search framework to evaluate alternative \ac{gNB} deployment configurations according to different optimization objectives. %
In particular, two placement criteria are investigated: coverage maximization and max-min signal optimization, the latter aiming to improve service conditions for users located in cell-edge regions. %
The proposed workflow allows the generation and evaluation of multiple deployment scenarios without requiring additional measurement campaigns. %


\section{Experimental Setup}



\begin{figure}[t]
		\centering
		\includegraphics[width=0.8\columnwidth]{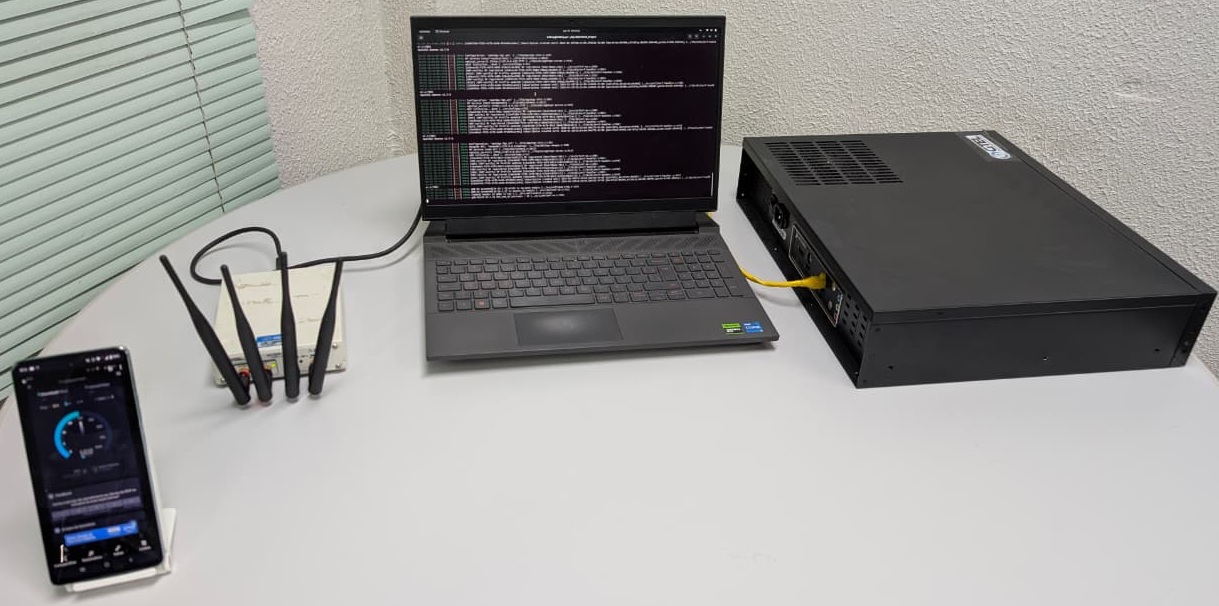}
		\caption{Experimental \ac{5G} testbed.}
		\label{Fig:Setup2}
\end{figure}

Our experimental \ac{5G} testbed is illustrated in~\FigRef{Fig:Setup2}. %
It was implemented as a standalone private \ac{5G} network composed of a \ac{gNB}, a \ac{5GC}, and a commercial \ac{UE}. %
The \ac{gNB} was implemented using the open-source srsRAN framework running on a laptop computer, which was connected through a USB interface to an Ettus Research USRP B210 acting as the radio front-end. %
The \ac{5GC} was implemented using Open5GS running on a desktop computer. %
The \ac{gNB} and the \ac{5GC} were interconnected through a wired Ethernet link, while the \ac{UE}, a Samsung Galaxy S25 smartphone, accessed the network through the \ac{5G} air interface. %
The Cellular-Pro software  was used to perform \ac{RSRP} measurements. %

The radio access link operated at $3.7$~GHz, corresponding to \ac{5G} \ac{NR} band~n78. %
This setup enabled the collection of \ac{RSRP} measurements from an experimental \ac{5G} deployment, capturing the effects of the radio front-end, protocol stack, commercial terminal, and indoor environment. %


\begin{figure}[t]
		\centering
		\includegraphics[width=\columnwidth]{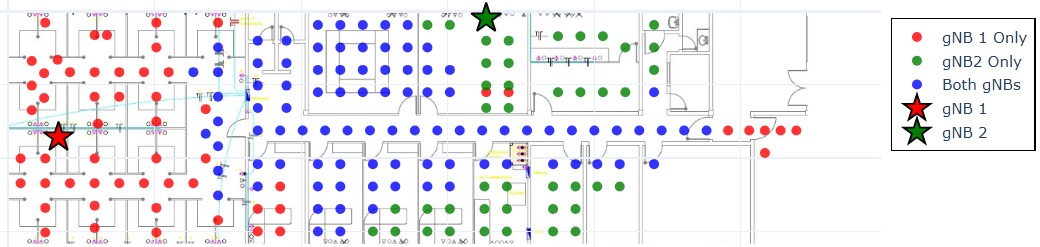}
		\caption{Indoor floor plan and \acs{gNB} locations.}
		\label{Fig:Planta-GTEL}
\end{figure}

Specifically, two measurement campaigns were conducted in an office environment. %
\FigRef{Fig:Planta-GTEL} shows the floor plan of the considered indoor scenario together with the measurement locations. %
The two stars identify the \ac{gNB} positions adopted during the campaigns. %
Blue dots correspond to locations measured in both campaigns, while red and green dots indicate points measured exclusively in the first and second campaigns, respectively. %
The environment spans approximately $35.64 \times 10.00$~m and comprises a variety of interconnected indoor spaces separated by multiple internal partitions, creating diverse propagation conditions that are representative of real-world office deployments. %



In the first campaign, the \ac{gNB} was positioned close to internal structural obstacles. %
This configuration intentionally created propagation conditions characterized by multiple wall crossings and severe attenuation, allowing the dataset to capture highly obstructed radio links typically found in indoor cellular deployments. %

In the second campaign, the \ac{gNB} was placed in a more open area of the floor plan with partial visibility towards the main corridor. %
This arrangement generated longer propagation paths with fewer obstacles and stronger line-of-sight components, complementing the propagation conditions observed in the first campaign. %

The combination of both datasets resulted in a diverse collection of \ac{RSRP} samples spanning different distances, visibility conditions, and obstruction levels. %
This diversity is essential for enabling the propagation model to generalize across distinct deployment configurations rather than specializing in a single \ac{gNB} location. %

Specifically, the first and second campaigns yielded 147 and 133 measurement points, respectively, resulting in a total dataset of 280~\ac{RSRP} samples. %
The measurements cover a wide range of \ac{gNB}--\ac{UE} distances and obstruction conditions, providing sufficient variability for training and evaluating the proposed propagation model. %

To support the subsequent propagation analysis, the environment was represented by a two-dimensional geometric model derived from the building floor plan. %
A total of $39$~wall segments were manually mapped and incorporated into the propagation analysis framework. %
This geometric representation enabled the automatic identification of structural obstacles intersecting the direct propagation path between the \ac{gNB} and \ac{UE} locations, allowing obstacle-related information to be incorporated into the learning process. %

\section{Data-Driven Propagation Modeling}

This section describes the methodology adopted to construct the propagation model used throughout the optimization process. %
The proposed approach combines feature engineering techniques with a machine-learning-based regressor trained from experimentally collected \ac{RSRP} measurements. %


The objective of the propagation model is to estimate the received signal level at arbitrary \ac{UE} locations while preserving the ability to generalize across different \ac{gNB} placements. %
A direct use of \ac{gNB} and \ac{UE} coordinates as model inputs would introduce a strong dependency on the specific geometries observed during training, potentially leading to spatial overfitting and poor generalization when evaluating new deployment configurations. %

To address this limitation, the original coordinate information was transformed into physically meaningful propagation descriptors. %
For each \ac{gNB}--\ac{UE} pair, two features were extracted: the Euclidean distance between the devices and the number of walls intersecting the direct propagation path. %

The \ac{gNB}--\ac{UE} distance is computed as
\begin{equation}
d_{i,k}
=
\sqrt{(X_i-X_k)^2+(Y_i-Y_k)^2},
\end{equation}
where $(X_i,Y_i)$ and $(X_k,Y_k)$ denote the coordinates of the \ac{UE} and \ac{gNB}, respectively.

The second feature quantifies the number of walls crossed by the direct propagation path connecting the \ac{gNB} and \ac{UE}. %
This parameter is obtained by evaluating the intersections between the line segment joining both locations and the wall segments defined in the floor plan:
\begin{equation}
w_{i,k}
=
\sum_{j=1}^{N_{\mathrm{walls}}}
\mathbb{I}
\Big(
\mathcal{L}(P_{TX_k},P_{RX_i})
\cap
\mathcal{W}_j
\neq
\emptyset
\Big),
\end{equation}
where $\mathbb{I}(\cdot)$ denotes the indicator function, $\mathcal{L}(\cdot)$ represents the direct propagation path, and $\mathcal{W}_j$ corresponds to the $j$-th wall segment in the geometric representation of the environment.

As a result, each measurement sample is represented by the feature vector
\begin{equation}
\mathbf{x}_{i,k}
=
[d_{i,k},w_{i,k}]^T,
\end{equation}
while the target variable corresponds to the measured \ac{RSRP} value associated with the considered \ac{gNB}--\ac{UE} pair. %

This representation enables the learning algorithm to capture the dominant attenuation mechanisms present in indoor propagation, namely distance-dependent path loss and obstacle penetration losses, while reducing the dependence on absolute \ac{gNB} coordinates.


The propagation model was implemented using the Light Gradient Boosting Machine (LightGBM)~\cite{ke2017lightgbm}, a gradient-boosting framework based on decision trees. %
LightGBM was selected due to its ability to efficiently model nonlinear relationships while maintaining low computational complexity and strong predictive performance on tabular datasets. %

Specifically, the model learns the nonlinear mapping $[d_{i,k},w_{i,k}] \rightarrow RSRP_{i,k}$ 
from the measurements collected during the experimental campaign. %

To evaluate its predictive performance, the dataset was randomly divided into training and validation subsets, comprising 70\% and 30\% of the available samples, respectively. %

Unlike analytical propagation models, which typically rely on predefined attenuation formulas, the adopted approach directly infers the relationship between environmental characteristics and received signal levels from empirical observations. %
This allows the model to capture propagation effects that may be difficult to represent analytically, including the combined influence of multiple obstacles and irregular indoor layouts.

Because the model relies exclusively on distance and obstacle information, it can estimate signal levels for previously unobserved \ac{gNB} locations, enabling the evaluation of hypothetical deployments without additional measurements.

\section{gNB Placement Optimization}

Once the propagation model is trained, it can be used to estimate the received signal level for arbitrary \ac{gNB} locations within the considered floor plan. %
This capability enables the evaluation of alternative deployment configurations without requiring additional measurement campaigns.

The objective of the optimization process is to determine the optimal placement of a set $\mathcal{K}$ of simultaneously deployed \acp{gNB} according to predefined performance criteria. %
For a given number of deployed \acp{gNB}, the optimization algorithm searches over the candidate location set and identifies the configuration that maximizes the selected objective function.

In this work, inter-cell interference is not explicitly considered. %
Instead, the deployed \acp{gNB} are assumed to operate on orthogonal frequency resources, such that the received signal quality at each location depends exclusively on the propagation conditions between the \ac{UE} and the serving \ac{gNB}. %
This assumption is equivalent to allocating distinct portions of the available spectrum to neighboring cells and allows the analysis to focus exclusively on the impact of \ac{gNB} placement on network coverage.

The optimization process is formulated as a combinatorial search problem in which a subset of candidate \ac{gNB} locations is selected according to a predefined performance criterion. %
For this, a set of candidate \ac{gNB} locations is uniformly distributed over the floor plan with a spatial resolution of 1~m. %
For each candidate position, the propagation model is used to estimate the \ac{RSRP} over the entire evaluation grid. %



When multiple \acp{gNB} are simultaneously deployed, the effective signal level experienced at each \ac{UE} location is assumed to correspond to the strongest available signal. %
Therefore, the effective received power at location $i$ is given by
\begin{equation}
RSRP_i^{\mathrm{eff}}
=
\max
\left\{
RSRP_{i,1},
RSRP_{i,2},
\dots,
RSRP_{i,|\mathcal{K}|}
\right\},
\label{eq:max_rsrp}
\end{equation}
where $\mathcal{K}$ denotes the set of active \acp{gNB} and $RSRP_{i,k}$ represents the signal level predicted for location $i$ from \ac{gNB} $k$.
This operation generates a composite coverage map representing the best signal available at each point of the environment. %


The first optimization criterion, called herein Coverage Maximization, seeks to maximize the percentage of locations where \ac{RSRP} exceeds a predefined coverage threshold. %
Let $\tau$ denote the minimum acceptable \ac{RSRP}. %
The optimal deployment is obtained by solving
\begin{equation}
\mathcal{K}^{*}_{\mathrm{cov}}
=
\arg\max_{\mathcal{K}\subset\mathcal{M}}
\frac{1}{|\mathcal{I}|}
\sum_{i\in\mathcal{I}}
\mathbb{I}
\left(
RSRP_i^{\mathrm{eff}}
\geq
\tau
\right),
\end{equation}
where $\mathcal{M}$ denotes the set of candidate locations and $\mathcal{I}$ represents the set of evaluation points.

In this work, the coverage threshold is fixed at $-95$~dBm. %
When multiple deployments achieve the same coverage percentage, the average \ac{RSRP} over the environment is used as a secondary selection criterion.


Although coverage maximization generally increases the served area, it may neglect locations experiencing poor propagation conditions. %
To improve service quality at the cell edge, a second optimization criterion is considered, called herein as Max--Min Optimization. %
Its objective is to maximize the weakest signal level observed throughout the environment:
\begin{equation}
\mathcal{K}^{*}_{\mathrm{maxmin}}
=
\arg\max_{\mathcal{K}\subset\mathcal{M}}
\left(
\min_{i\in\mathcal{I}}
RSRP_i^{\mathrm{eff}}
\right).
\end{equation}
It prioritizes deployments that improve the worst-case propagation conditions, reducing coverage holes and enhancing signal availability in challenging regions of the floor plan.

Similarly to the other criterion, ties are resolved using the average effective \ac{RSRP} as a secondary metric.

\section{Results and Discussion}


First, we evaluated the predictive performance of the adopted propagation model, i.e., LightGBM, through a comparison with alternative machine-learning regressors, namely \ac{KNN}, \ac{ADA}, \ac{GBR}, and \ac{RF}. %
Model accuracy was assessed using the \ac{MAE}, \ac{MSE}, and \ac{RMSE}. %
\TabRef{tab:comparativo_modelos} summarizes the obtained results.

\begin{table}[t]
\caption{Prediction performance of the evaluated regression models}
\label{tab:comparativo_modelos}
\centering
\begin{tabular}{lccc}
\toprule
\textbf{Regressor} & \textbf{MAE (dB)} & \textbf{MSE (dB$^2$)} & \textbf{RMSE (dB)} \\
\midrule
\textbf{LightGBM}  & \textbf{5.50} & \textbf{51.02} & \textbf{7.02} \\
\acs{KNN}          & 5.27          & 51.70          & 7.03          \\
\acs{ADA}          & 5.83          & 55.18          & 7.31          \\
\acs{GBR}          & 5.79          & 69.12          & 8.12          \\
\acs{RF}           & 5.96          & 69.16          & 8.15          \\
\bottomrule
\end{tabular}
\end{table}

The results indicate that LightGBM achieved the best overall predictive performance among the evaluated models, yielding the lowest \ac{RMSE} and \ac{MSE} values. %
Although \ac{KNN} obtained a slightly lower \ac{MAE}, LightGBM exhibited greater robustness against larger prediction errors, resulting in superior overall accuracy. %

Although the obtained prediction errors may appear relatively high at first glance, they should be interpreted in the context of the considered deployment scenario. %
Indoor radio propagation at $3.7$~GHz is strongly affected by multipath propagation, shadowing, furniture, wall composition, and other environmental factors that are not explicitly represented by the adopted feature set. %
Despite relying exclusively on two physically interpretable features, namely \ac{gNB}--\ac{UE} distance and wall count, the proposed model achieved an \ac{RMSE} close to $7$~dB, which is comparable to the variability commonly observed in indoor propagation environments~\cite[Table 7.5-6 Part-1: Indoor Office]{3gpp.38.901d}. %
These results indicate that the selected feature representation captures the dominant propagation mechanisms with sufficient accuracy to support deployment planning and comparative placement analysis.
Consequently, LightGBM was selected as the propagation model employed throughout the subsequent placement optimization analysis. %



\begin{figure}[!t]
	\centering
	
	\begin{subfigure}{\columnwidth}
		\centering
		\includegraphics[trim={1.3cm 3cm 1cm 4,08cm}, clip,width=1.0\columnwidth]{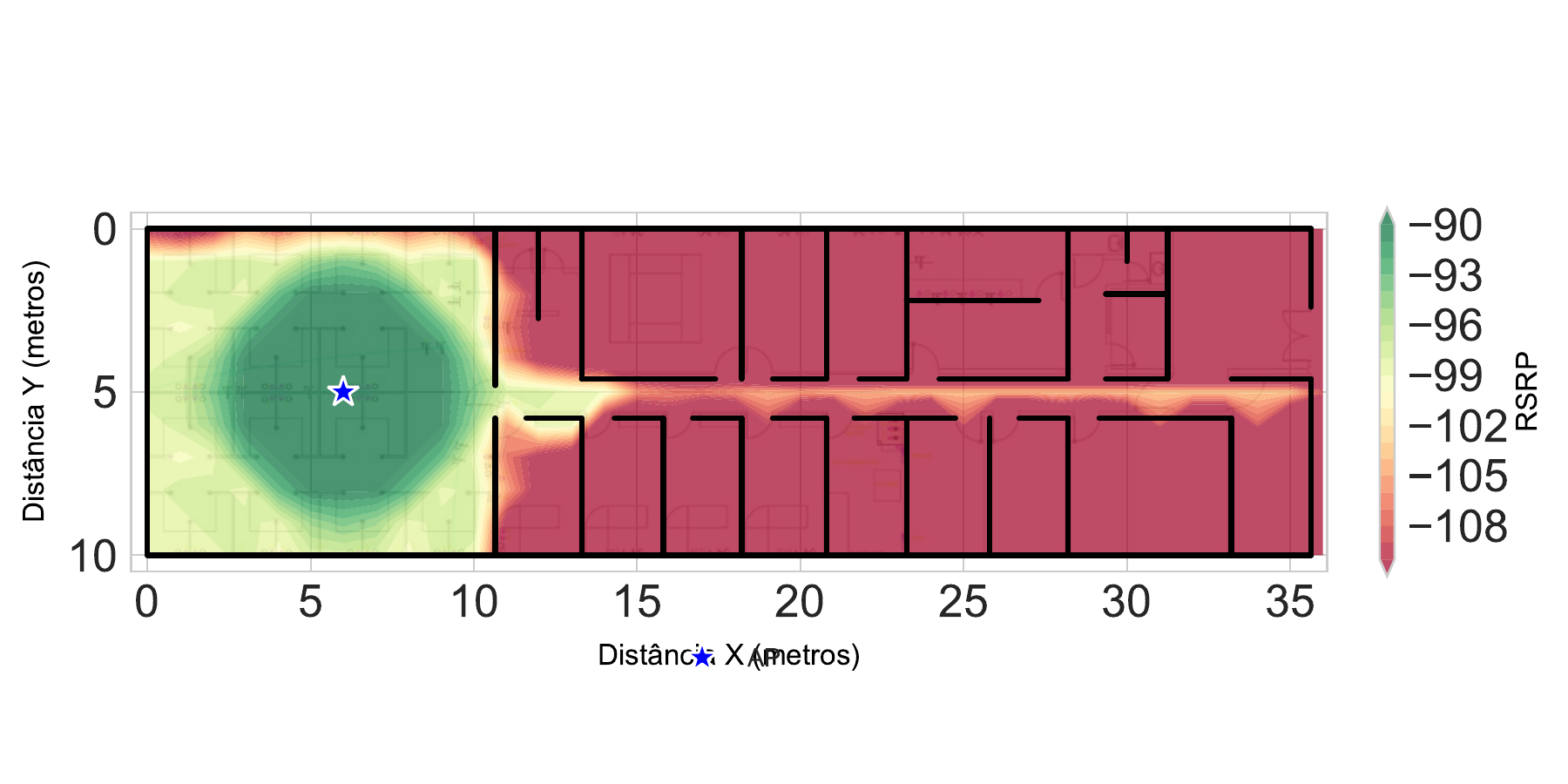}
		\caption{Coverage Maximization (1 gNB).}
		\label{fig:1gnb_cov}
	\end{subfigure}
	
	\vspace{0.3cm}
	
	\begin{subfigure}{\columnwidth}
		\centering
		\includegraphics[trim={1.3cm 3cm 1cm 4,08cm}, clip,width=1.0\columnwidth]{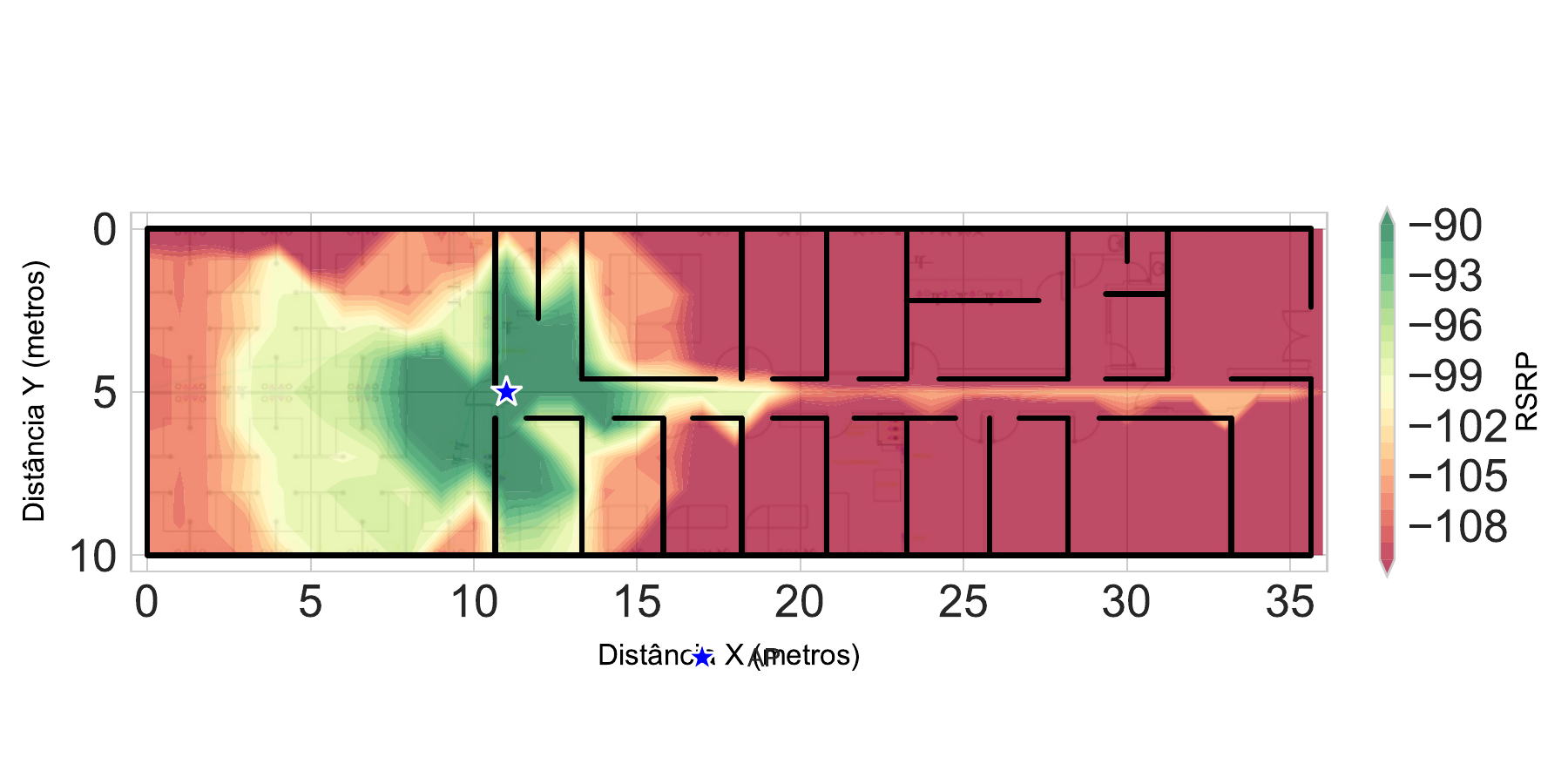}
		\caption{Max--Min Optimization (1 gNB).}
		\label{fig:1gnb_maxmin}
	\end{subfigure}
	
	\caption{\label{fig:comparativo_mapas_1gnb}
		Coverage maps obtained for a single deployed gNB under the two optimization criteria.}
\end{figure}

\begin{figure}[!t]
	\centering
	
	\begin{subfigure}{\columnwidth}
		\centering
		\includegraphics[trim={1.3cm 3cm 1cm 4,08cm}, clip,width=1.0\columnwidth]{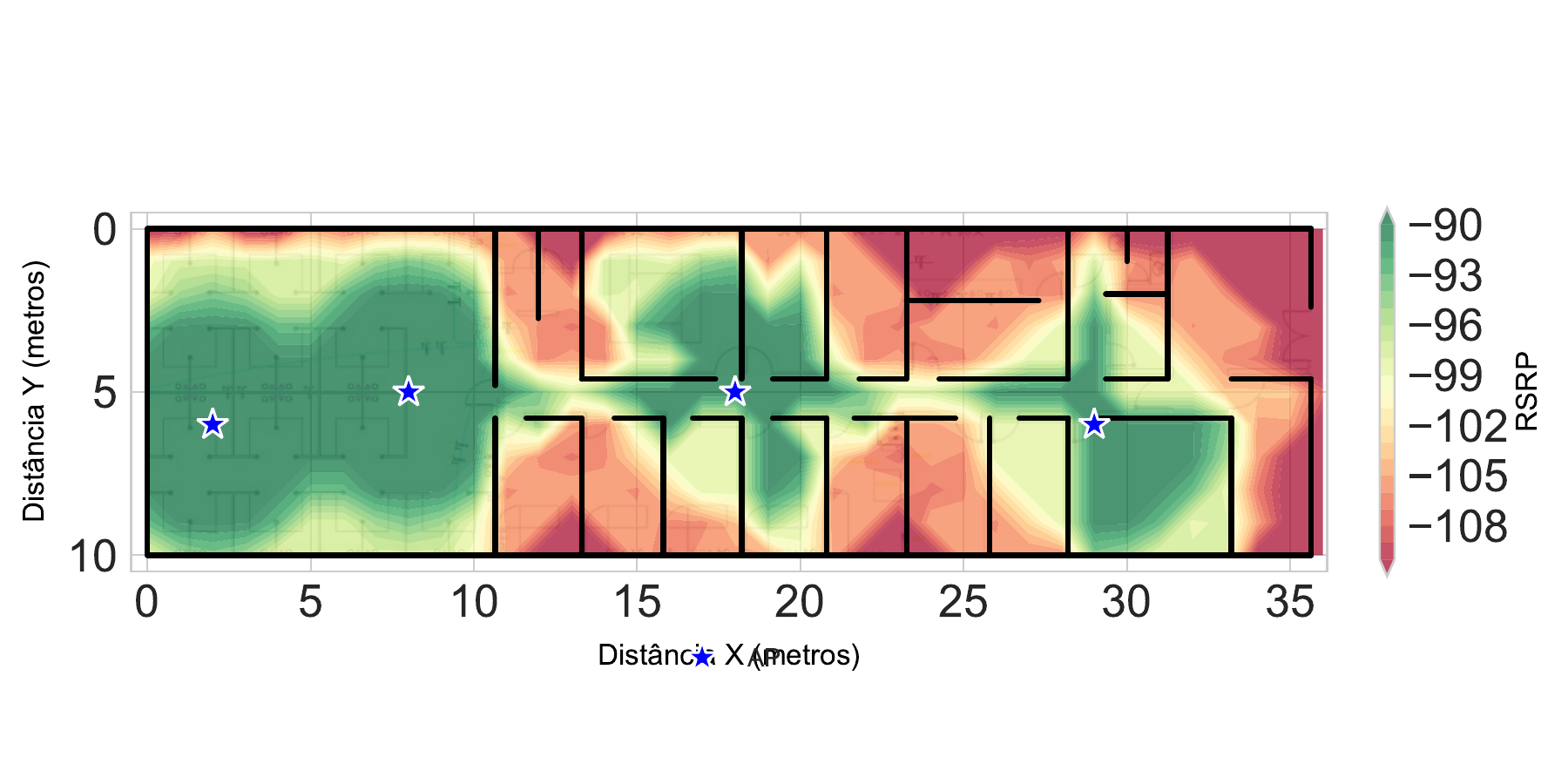}
		\caption{Coverage Maximization (4 gNBs).}
		\label{fig:4gnb_cov}
	\end{subfigure}
	
	\vspace{0.3cm}
	
	\begin{subfigure}{\columnwidth}
		\centering
		\includegraphics[trim={1.3cm 3cm 1cm 4,08cm}, clip,width=1.0\columnwidth]{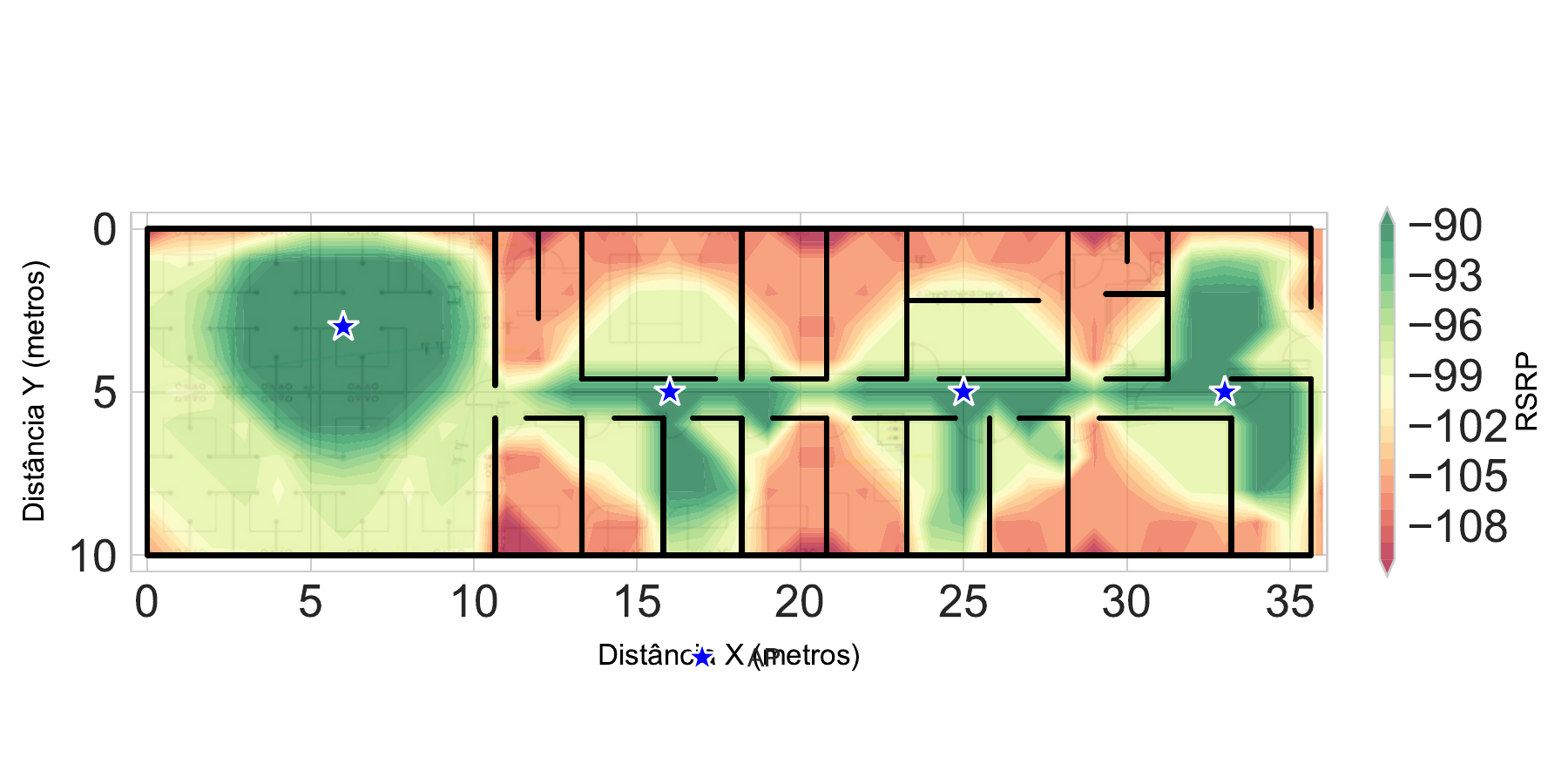}
		\caption{Max--Min Optimization (4 gNBs).}
		\label{fig:4gnb_maxmin}
	\end{subfigure}
	
	\caption{\label{fig:comparativo_mapas_4gnb}
		Coverage maps obtained for four simultaneously deployed gNBs under the two optimization criteria.}
\end{figure}

Figures~\ref{fig:comparativo_mapas_1gnb} and~\ref{fig:comparativo_mapas_4gnb} illustrate the coverage maps obtained under the coverage-maximization and max--min criteria for the one- and four-\ac{gNB} deployment scenarios, respectively. %

It should be emphasized that the evaluated multi-\ac{gNB} scenarios are not intended to represent realistic deployment recommendations for the considered office environment. %
In fact, given the relatively small dimensions of the investigated floor plan, deploying several closely spaced \acp{gNB} would rarely be justified in practice from either a coverage or cost perspective. %
Instead, these scenarios are used to investigate how different optimization objectives influence \ac{gNB} placement and coverage characteristics as additional deployment degrees of freedom become available. %
The considered configurations should therefore be interpreted as a controlled sensitivity analysis that provides a broader validation of the proposed optimization framework under increasingly flexible deployment conditions. %
Furthermore, it is important to highlight that, although the selected office environment represents a relatively small-scale scenario, the proposed methodology is not restricted to this setting and can be readily extended to larger indoor deployments, where multiple \acp{gNB} are often required to satisfy coverage and capacity requirements. %

For the single-\ac{gNB} case (\FigRef{fig:comparativo_mapas_1gnb}), the criteria place the \ac{gNB} in different regions of the floor plan. %
The coverage-maximization solution selects a location near the left side of the building, where a large open area can be reached with relatively limited wall penetration. %
In contrast, the max--min formulation positions the \ac{gNB} closer to the offices area, reducing the maximum distance to remote locations and improving signal availability in the corridor and offices located further from the \ac{gNB}. %

As a consequence, the coverage-oriented solution provides stronger signal levels over a larger contiguous area, whereas the max--min placement produces a more balanced signal distribution throughout the environment. %
Although neither configuration is capable of providing satisfactory coverage to the entire floor plan with a single \ac{gNB}, the difference between the optimization philosophies already becomes apparent.

The distinction between the optimization objectives becomes considerably clearer when four \acp{gNB} are available. %
Under the coverage-maximization criterion, the \acp{gNB} are distributed so as to expand the area exceeding the target \ac{RSRP} threshold, creating several regions with strong signal levels while leaving some highly attenuated zones largely unchanged. %
Conversely, the max--min solution places the \acp{gNB} in a manner that specifically targets poorly covered regions, resulting in a noticeably more homogeneous coverage pattern across the entire floor plan.

This behavior illustrates the fundamental trade-off between the two optimization objectives. %
The coverage-maximization criterion seeks to increase the number of adequately served locations, whereas the max--min formulation prioritizes improvements in the most challenging propagation conditions. %
The visual differences observed in the coverage maps suggest that the latter strategy may be particularly attractive for industrial and enterprise deployments where service continuity across the entire operational area is more important than maximizing coverage in already well-served regions.

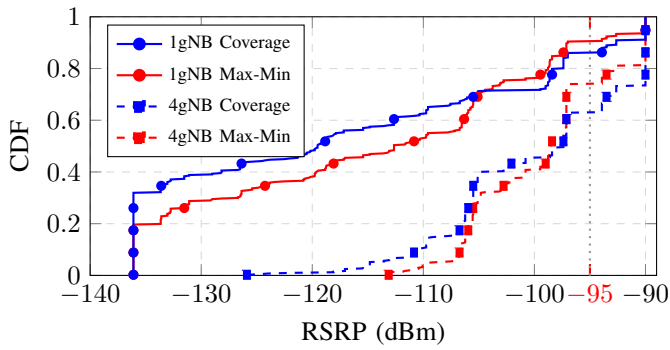
\begin{figure}[htbp]
    \centering
    \begin{tikzpicture}
        \begin{axis}[
            xlabel={RSRP (dBm)},
            ylabel={CDF},
            xmin=-140, xmax=-89,   
            ymin=0, ymax=1,
            grid=major,            
            grid style={dashed, gray!30},
            legend pos=north west,  
            legend style={
                font=\scriptsize, 
                row sep=1.5pt, 
                legend columns=1,
            }, 
            width=0.5\textwidth,   
            height=5cm,            
            extra x ticks={-95},
            extra x tick style={
                tick label style={color=red, font=\bfseries, anchor=north},
                major tick style={color=red, thick},
                grid=none
            }
        ]
        
        \addplot[gray!80, thick, dotted, forget plot] coordinates {(-95, 0) (-95, 1.12)};

        \addplot[color=red, thick, mark=*, mark repeat=35, mark size=1.5pt, forget plot] 
            table [x index=2, y index=0, col sep=comma] {Figs/CDF-DADOS.csv};

        \addplot[color=blue, thick, mark=*, mark repeat=35, mark size=1.5pt, forget plot] 
            table [x index=1, y index=0, col sep=comma] {Figs/CDF-DADOS.csv};        
            
        \addplot[color=red, thick, dashed, mark=square*, mark repeat=35, mark size=1.5pt, forget plot] 
            table [x index=4, y index=0, col sep=comma] {Figs/CDF-DADOS.csv};

        \addplot[color=blue, thick, dashed, mark=square*, mark repeat=35, mark size=1.5pt, forget plot] 
            table [x index=3, y index=0, col sep=comma] {Figs/CDF-DADOS.csv};

        
        \addlegendimage{color=blue, thick, mark=*}
        \addlegendentry{1gNB Coverage}
        
        \addlegendimage{color=red, thick, mark=*}
        \addlegendentry{1gNB Max-Min}
        
        \addlegendimage{color=blue, thick, dashed, mark=square*}
        \addlegendentry{4gNB Coverage}
        
        \addlegendimage{color=red, thick, dashed, mark=square*}
        \addlegendentry{4gNB Max-Min}
%
%
%
%
        
        \end{axis}
    \end{tikzpicture}
    \caption{\acs{CDF} for the one- and four-\ac{gNB} deployment scenarios under the coverage maximization and max--min optimization criteria, highlighting the 10th percentile.}
	\label{fig:cdf_comparativa}
\end{figure}

The cumulative effect of these deployment decisions can be further observed in the \acp{CDF} shown in Fig.~\ref{fig:cdf_comparativa}. %

For the single-\ac{gNB} scenario, both optimization criteria produce very similar lower-tail behavior, reflecting the limited flexibility available when only one \ac{gNB} can be deployed. %
However, differences emerge in the four-\ac{gNB} case. %
The max--min formulation shifts the lower tail of the distribution toward higher \ac{RSRP} values, indicating an improvement for users located in unfavorable propagation conditions. %
This effect is particularly visible around the highlighted 10th percentile, which captures the behavior of the weakest covered locations.

While the coverage-maximization criterion maintains a larger fraction of locations above the target threshold, the max--min strategy reduces the severity of coverage holes and improves the worst service conditions throughout the environment. %
Therefore, the \acp{CDF} confirm that the observed differences are not restricted to isolated regions of the floor plan but instead affect the overall spatial distribution of received signal levels.

To generalize these observations, Table~\ref{tab:otimizacao_resumo} summarizes the performance obtained for all deployment scenarios ranging from one to four simultaneously active \acp{gNB}.

\begin{table}[hbt]
	\centering
	\caption{Optimization Criteria Comparison}
	\label{tab:otimizacao_resumo}
	\resizebox{\columnwidth}{!}{%
		\begin{tabular}{lcccc}
			\toprule
			&
			$\mathbf{N_{\mathrm{gNB}}}$ &
			\textbf{\shortstack{Coverage \\ Ratio (\%)}} &
			\textbf{\shortstack{10th Percentile \\ (dBm)}} &
			\textbf{\shortstack{Worst-Case \\ RSRP (dBm)}} \\
			\midrule
			
			\multirow{4}{*}{\rotatebox[origin=c]{90}{\textbf{Coverage}}}
			& 1 & 14.0 & -131.7 & -136.1 \\
			& 2 & 22.4 & -130.7 & -131.1 \\
			& 3 & 30.5 & -130.7 & -130.7 \\
			& 4 & 37.1 & -111.5 & -120.7 \\
			
			\midrule
			
			\multirow{4}{*}{\rotatebox[origin=c]{90}{\textbf{Max--Min}}}
			& 1 & 9.6 & -131.1 & -136.1 \\
			& 2 & 15.5 & -116.5 & -120.7 \\
			& 3 & 22.4 & -107.5 & -116.5 \\
			& 4 & 26.0 & -106.3 & -113.2 \\
			
			\bottomrule
		\end{tabular}%
	}
\end{table}

The quantitative results corroborate the trends previously observed in both the coverage maps and the \acp{CDF}. %
For all evaluated deployment scenarios, the coverage-maximization formulation achieves the largest coverage ratio, whereas the max--min criterion consistently provides superior performance for both the 10th-percentile and worst-case \ac{RSRP} metrics.

More importantly, the results indicate that the benefits of the max--min strategy become increasingly pronounced as additional deployment resources become available. %
From two deployed \acp{gNB} onward, substantial improvements are observed in the lower-tail metrics, demonstrating that the optimization process successfully exploits the additional placement flexibility to improve the most disadvantaged locations. %
At the same time, the coverage-maximization criterion continues to favor a larger served area, producing a persistent trade-off between coverage expansion and cell-edge performance.

Overall, the results demonstrate that the proposed data-driven framework can support indoor radio planning by identifying deployment configurations tailored to different operational objectives. %
Depending on the target application, network designers may prioritize either maximizing the covered area or improving service reliability in the most challenging regions of the environment, both of which can be systematically evaluated using the proposed methodology.

\section{Conclusion}

This paper presented a data-driven framework for indoor \ac{gNB} placement optimization based on measurements collected from a \ac{5G} testbed. %
A LightGBM-based propagation model was trained from experimental \ac{RSRP} measurements and used to evaluate alternative deployment configurations under coverage-maximization and max--min optimization criteria.

Results demonstrated that the proposed methodology can support indoor radio planning while revealing the trade-off between coverage expansion and cell-edge performance. %
Future work includes optimization formulations incorporating interference and capacity constraints.

%
%



\printbibliography

@IEEEtranBSTCTL{IEEEexample:BSTcontrol,
  CTLuse_forced_etal       = "yes",
  CTLmax_names_forced_etal = "2",
  CTLnames_show_etal       = "1" 
}

@String { std3GPP          = {3rd Generation Partnership Project {(3GPP)}} }

@TechReport{3gpp.38.901d,
	Title                    = {Study on Channel Model for Frequencies from 0.5 to 100 {GHz}},
	Author                   = {3GPP},
	%Institution              = std3GPP,
	Year                     = {2026},
	Month                    = mar,
	Number                   = {{38.901}},
	Type                     = {TR},
	Note                     = {v.18.3.0},
	%Url                      = {http://www.3gpp.org/DynaReport/38901.htm},
	%Urldate		   		   = {2026-05-22}
}

@techreport{ITU_R_P1238_13_2025,
  author       = {{ITU}},%{{International Telecommunication Union}},
  title        = {Propagation Data and Prediction Methods for the Planning of Indoor Radiocommunication Systems and Radio Local Area Networks in the Frequency Range from 300 {MHz} to 450 {GHz}},
  institution  = {ITU Radiocommunication Sector (ITU-R)},
  type         = {ITU-R},
  number       = {P.1238-13},
  year         = {2025},
  month        = sep,
  address      = {Geneva, Switzerland},
  %url          = {https://www.itu.int/rec/R-REC-P.1238},
  %note         = {Accessed: May 18, 2026}
}

@techreport{Ericsson_Mobility_Indoor_2023,
  author      = {{Ericsson}},
  title       = {Demand for Indoor Connectivity Driving the Need for Enhanced Performance},
  institution = {Ericsson},
  type        = {Ericsson Mobility Report},
  year        = {2023},
  month       = nov,
  %url         = {https://www.ericsson.com/4ad0e9/assets/local/reports-papers/mobility-report/documents/2023/emr-november-2023-indoor-connectivity-article.pdf},
  %note        = {Accessed: May 18, 2026}
}

@misc{alves2024,
      title={Experimental comparison of {5G} {SDR} platforms: {srsRAN} x {OpenAirInterface}}, 
      author={Alves, Ruan P. and Joao Guilherme A. da S. Alves and Mikael R. Camelo and Wilker O. de Feitosa and Victor F. Monteiro and Fco. Rodrigo P. Cavalcanti},
      year={2024},
      eprint={2406.01485},
      archivePrefix={arXiv},
      %primaryClass={cs.NI},
      %url={https://arxiv.org/abs/2406.01485}, 
}

@article{ke2017lightgbm,
  title={Lightgbm: A highly efficient gradient boosting decision tree},
  author={Ke, Guolin and Meng, Qi and Finley, Thomas and Wang, Taifeng and Chen, Wei and Ma, Weidong and Ye, Qiwei and Liu, Tie-Yan},
  journal={Advances in neural information processing systems},
  volume={30},
  year={2017}
}


\end{document}